\documentclass[aps,prc,twocolumn,superscriptaddress,showpacs,showkeys]{revtex4}
\usepackage{amsmath,amssymb}
\usepackage{bm}
\usepackage[dvips]{graphicx}
\input{epsf}
\newcommand{\be}{\begin{eqnarray}}
\newcommand{\ee}{\end{eqnarray}}

\newcommand{\ket}{\rangle}
\newcommand{\bra}{\langle}

\begin{document}

\title{Structure and production of $\Theta^+$}

\author{A. Hosaka}%
\affiliation{Research Center for Nuclear Physics (RCNP), 
Ibaraki, Osaka 567-0047, Japan}%


\begin{abstract}
    We study properties of the pentaquark particle $\Theta^+$ 
	with emphasis on the role of chiral symmetry.  
	It is shown that when chiral force is sufficiently strong, the 
	positive parity $\Theta^+$ may be realized with a lower 
	mass relative to the negative parity state.
	The decay width is then studied in the non-relativistic  
	quark model.  
	It is shown that the narrow width may be realized for the 
	positive parity state, while the seemingly lowest negative parity 
	state couples strongly to the continuum state resulting in a 
	very broad width.  
	Finally, $\Theta^+$ production is studied 
	in photo-induced and proton-induced processes.  
	The polarized proton reaction provides a model independent 
	method to determine the parity of $\Theta^+$.   
\end{abstract}


\maketitle

\section{Introduction}

The observation of evidence of the pentaquark particle 
$\Theta^+$ has triggered 
enormous amount of works both in experiments and 
theories~\cite{Nakano:2003qx}. 
The investigation requires new insight to 
understand the properties of the multiquark system with 
at least five quarks~\cite{oka}.  
It is a great challenge to explore from the 
conventional hadronic matter to a new type of  
matter eventually toward the quark matter.  
It is also expected that the study of the multiquark system will 
provide information for non-perturbative QCD at low 
energies, which we have not yet understood well. 
The mechanism for the light mass and narrow width of $\Theta^+$ 
is obviously an important issue to be solved 
together with the determination of its spin and parity.  
In particular, the parity reflects the internal motion of the 
state and should be a crucial observable for the 
development of the new field as described above.   

In the previous studies of hadrons, 
the constituent quark model and the mesonic soliton model 
have been often used as typical QCD oriented 
models~\cite{quark,skyrme}.  
The basic assumptions of these models look rather different,  
but various properties of hadrons, especially of baryons, have been 
reproduced in a similar manner once several
fundamental parameters in the models are fixed appropriately.  
This is because many properties of the ground
state hadrons are dictated mostly by the symmetry, especially
the flavor symmetry of QCD.  
If we go to higher energy states, however, one expects to 
observe dynamical effects of the interaction, 
which should reflect the difference in various models.  
Therefore,  it is of great importance 
to explore the properties of multiquark 
systems in various models.   

In this report, first we discuss the role of chiral symmetry 
on the parity and mass of the pentaquark particle.  
Chiral symmetry is the fundamental symmetry of QCD and its 
spontaneous breaking governs the dynamics of low energy 
hadrons.  
As we will see, the study of multiquark systems reveals 
another interesting features of the dynamics of chiral symmetry.  
In fact, the success of the original work in the chiral soliton 
model has already implied its crucial role~\cite{Diakonov:1997mm}.
To see this point explicitly, we adopt the 
chiral bag model which accommodates quark single particle levels 
depending on the strength of the pion-quark interaction at the 
bag surface~\cite{cbag}.  
The pion quark interaction induces level crossing between 
the positive and negative parity states, giving an intuitive 
way to understand the origin of the parity of 
$\Theta^+$~\cite{Hosaka:2003jv}.  
The result of the chiral soliton 
model of positive parity $\Theta^+$ and that of the naive 
quark model of negative parity may be explained 
depending on the strength of the pion-quark interaction.  

As a physical quantity which reflects the parity, 
we investigate the strong decay of $\Theta^+$.  
For actual computation, we adopt the quark model and consider 
effects of different configurations of quark 
single particle wave functions of 
the spin, flavor, color and orbital spaces.  
The direct role of chiral symmetry for the decay amplitude
itself should be studied separately, which we 
do not consider here.  

In the second part we discuss production reactions
of $\Theta^+$.  
Photo-induced and polarized proton induced reactions 
are investigated, where the total and angular distributions 
of the cross sections are studied.  
In particular, the latter provides a selection rule to 
determine the parity in a model independent manner.

\section{Role of chiral symmetry for $\Theta^+$}

Let us start with the quark model, which has been 
successfully applied to the description of conventional hadrons 
and should be tested also for the pentaquark state~\cite{jennings}.  
In a naive constituent quark model, valance quarks 
occupy single particle states which are given by 
the harmonic oscillator.  
Due to the degeneracy of the spin-flavor and color degrees of 
freedom, for the ground state of the pentaquark $\Theta^+$, 
four quarks of $uudd$ and one antiquark $\bar s$ occupy the 
lowest $0s$ state, hence the configuration of $(0s)^5$.  
In the constituent quark model, the mass of the ground state 
is given by the sum of the five constituent quark masses.   
This amounts to be about 1.7 GeV when 
$m_{u, d} \sim 310$ MeV and $m_{s} \sim 500$ MeV are employed.  
The value 1.7 GeV is slightly larger than 
the experimental value $\sim$ 1.5 GeV.  
The mechanism has been proposed to reduce the mass by the 
spin-flavor or spin-color interaction as mediated by the 
Nambu-Goldstone meson exchange or gluon exchange~\cite{oka}.   

The parity of the $(0s)^5$ configuration is negative, because 
the parity of $\bar s$ is negative.  
Therefore, one would expect that the naive quark model 
would predict the negative parity state for the lowest 
pentaquarks.  
As will be shown later, however, such a $(0s)^5$ configuration 
couples strongly to the $KN$ continuum, resulting in 
a very wide decay width. 
In contrast, for positive parity state the decay width 
can be of order of 10 MeV.  

If we make a positive parity pentaquark state in the quark model, 
one of the quarks must be excited into a p-wave ($l=1$) orbit.  
This costs another 500 MeV for the mass of $\Theta^+$.  
A relevant question is if there is a mechanism to lower such 
a high lying state as compared with the negative parity state.  
As shown in Ref.~\cite{stancu} the flavor dependent 
force due to the Nambu-Goldstone boson exchanges between quarks 
has a large negative matrix element for the most attractive 
channel of the pentaquark state.  

\begin{figure}[tbp] 
\epsfxsize=4cm
\epsfbox{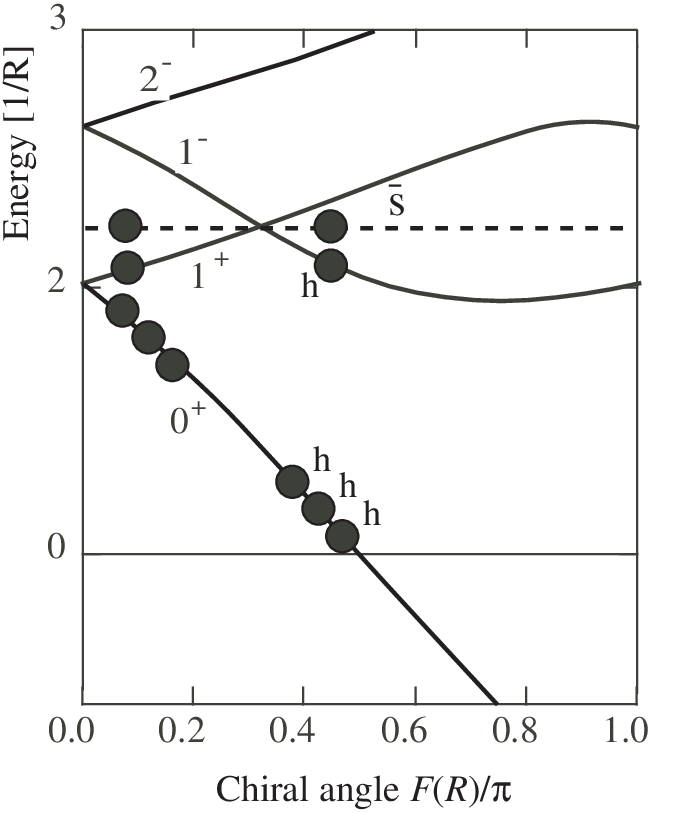}
\caption{Quark energy levels of the chiral bag model in the 
hedgehog configuration as functions of the chiral angle $F$. 
\label{cbag_level}}
\end{figure}

In the chiral bag model, the change in the parity can be 
also illustrated
by plotting quark single particle states as functions 
of the chiral angle at the bag surface 
(see Fig.~\ref{cbag_level})~\cite{Hosaka:2003jv}. 
In the hedgehog configuration, the chiral invariant pion-quark
interaction reduces to a spin-isospin interaction 
of the type $\vec \sigma \cdot \vec \tau$
which resolves the degeneracy in the 
$K^P = (j\pm 1/2)^P$ states, with $K = j - 1/2$ state  
lowered and with $K = j + 1/2$ state pushed up.  
Here $K$ is the sum of the spin and isospin $K = J+I$~\cite{cbag}.  
As a consequence, for sufficiently strong pion-quark interaction, 
the $K^P = 1^-$ level becomes lower than the $K^P = 0^+$ level, 
where the $(0s)^4(0p)^1$ configuration is realized, replacing the 
$(0s)^5$ configuration as the lowest energy state.  
Hence the positive parity state could be the lowest energy configuration.  
In the chiral bag model for the nucleon, the bag radius of 
about $R \sim 0.6$ (chiral angle $F \sim \pi/2$) is 
favored~\cite{cbag}.  
If this would also be the case for the pentaquark, 
the mass of the positive parity state is slightly 
lower than the negative parity one $\sim 1.7$ GeV.  

\vspace*{0.5cm}

\section{Decay width}

\vspace*{0.5cm}
\begin{figure}[tbp] 
\epsfxsize=7cm
\epsfbox{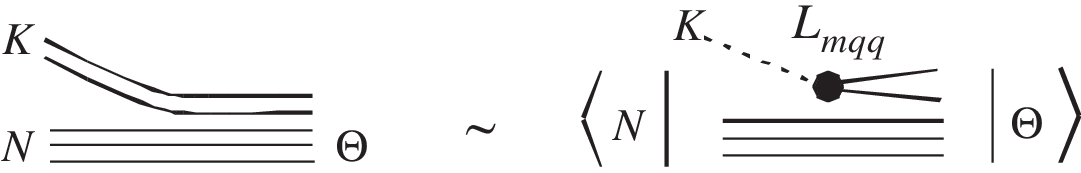}
\caption{Decay of the pentaquark state.  
\label{fallapart}}
\end{figure}

The decay of the pentaquark state going into one baryon and 
one meson is described by the fall-apart process as shown in 
Fig.~\ref{fallapart} (left).  
The matrix element of such a process is written as a product of the 
so-called spectroscopic factor and an interaction 
matrix element, 
\be
{\cal M}_{\Theta^+ \to KN} = 
S_{KN \; {\rm in} \; \Theta^+} \cdot h_{int} \, .
\ee
The former 
$S_{KN \; {\rm in} \; \Theta^+}$
is an amplitude to find in the pentaquark state
three-quark and quark-antiquark clusters having the 
quantum numbers of the nucleon and kaon, 
respectively.  
Explicit calculation for this factor was 
done in Ref.~\cite{carlson}.  
In the quark model, the interaction matrix element
may be computed by the meson-quark interaction of Yukawa type:
\be
L_{mqq} = g \bar q \gamma_5 \lambda_a \phi^a q\, , 
\ee
where $\lambda_a$ are SU(3) flavor matrices and 
$\phi^a$ are the octet meson fields.  
The coupling constant $g$ may be determined from the 
pion-nucleon coupling constant $g_{\pi NN} = 5g$.  
Therefore, using $g_{\pi NN} \sim 13$, we find $g \sim 2.6$.

The transition $\Theta^+ \to KN$ contains a matrix 
element of quark-antiquark annihilation of, for instance, 
$\bra 0 | L_{mqq} | u \bar s\ket$ as shown 
in Fig.~\ref{fallapart} (right).  
Details of calculation will be reported elsewhere~\cite{hosaka2}, 
and here several results are summarized as follows.  
For the negative parity state of $(0s)^5$, the decay 
width turns out to be of order of several hundreds MeV, 
typically 0.5 $\sim$ 1 GeV.  
In the calculation it has been assumed that the spatial wave function 
for the initial and final state hadrons are described by the 
common harmonic oscillator states.  
Also the masses of the particles are taken as experimental 
values, e.g., $M_{\Theta^+} = 1540$ MeV.  
For the result of the negative parity state of $(0s)^5$, 
the unique prediction can be made, since there is only one 
quark model states.  
The very broad width suggests that the $(0s)^5$ state 
couples very strongly to the $KN$ continuum and 
is hardly identified with a resonant state with 
a narrow width.  

The computation for the positive parity state is 
slightly complicated, since the orbital excitation 
introduces additional degree of freedom when writing the 
quark model wave function. 
In fact, four independent configurations 
are available for spin-parity $J^P = 1/2^+$~\cite{jennings}.  
Here we consider three configurations which minimize 
(1) a spin-flavor interaction of  one meson 
exchange~\cite{carlson}, 
(2) a spin-color interaction of  one gluon 
exchange, and 
(3) the $S=I=0$ diquark correlated state as proposed 
by Jaffe and Wilczek~\cite{Jaffe:2003sg}.  
The resulting decay widths are about
80 MeV, 40 MeV and 10 MeV, respectively.  
The diquark correlation of (3) develops a spin-flavor-color 
wave function having small overlap with the decaying channel of 
the nucleon and kaon.   
The decay width may be further suppressed by including 
correlations of spatial wave functions.  
Therefore, in the quark model, it is possible to 
reproduce the experimentally observed 
narrow width of $\Theta^+$ for the positive parity.  

The small values of the decay width for $J^P = 1/2^+$ 
as compared with the large values for $J^P = 1/2^-$ may be
explained by the difference in the coupling structure; 
one is the pseudoscalar type of $\vec \sigma \cdot \vec q$
and the other the scalar type of 1.  
The former of the p-wave coupling 
includes a factor $q/(2M)$ which suppresses the 
decay width significantly as compared with the latter
at the present kinematics, $q \sim 250$ MeV and 
$M \sim 1$ GeV, when the same coupling constant 
$g_{NK\Theta}$ is employed.

\section{Production of $\Theta^+$}

The $\Theta^+$ production from the 
non-strange initial hadrons is furnished by 
the creation of $s \bar s$ pair, which requires 
energy deposit around 1 GeV.  
In general the reaction mechanism of such energy 
region is considered to be complicated.  
However, as one of calculable methods, we adopt an 
effective lagrangian approach.  
Input parameters in the lagrangians should reflect 
the properties of $\Theta^+$ and therefore, 
the comparison of calculations and experiments will help
study the structure of $\Theta^+$.  
One of the most important ones is, as 
discussed in the previous section, the parity.  
Here we briefly discuss (1) photoproduction as
originally performed in LEPS 
group~\cite{Nam:2003uf,refphoto} 
and 
(2) $\Theta^+ \Sigma^+$ production induced by 
the polarized $\vec p \vec p$.  

\subsection{Photoproduction}

\begin{figure}[tbp] 
\epsfxsize=8cm
\epsfbox{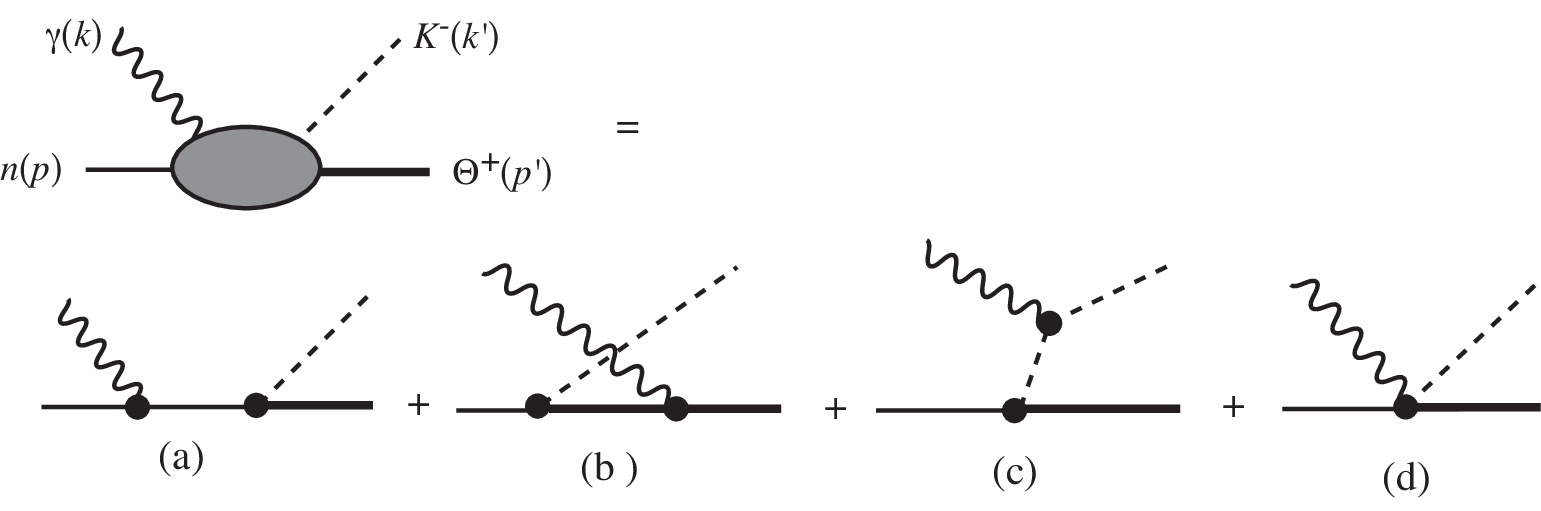}
\caption{Born diagrams for the $\Theta^+$ photoproduction. 
\label{photo_tree}}
\end{figure}

As described in detail in Ref.~\cite{Nam:2003uf}, 
in the effective lagrangian method we calculate the 
Born (tree) diagrams as depicted in Fig.~\ref{photo_tree}.  
The actual form of the interaction lagrangian depends on the 
scheme of introduction, i.e., either pseudoscalar 
(PS) or pseudovector (PV) schemes.  
In the PS, the three Born diagrams (a)-(c) are 
computed with gauge symmetry maintained.  
In the PV, on the contrary, the contact Kroll-Ruderman 
term (d) is also necessary.  
In the PS scheme, the contact term may be included in the 
antinucleon contribution of the nucleon pole terms.  
If chiral symmetry is respected, the low energy theorems 
guarantee that the two schemes should provide the same answer.  
In reality, due to the large energy deposit of order 
1 GeV, one worries that the equivalence may be violated.  
It is shown that the difference in the two schemes 
is proportional to the kaon momentum in the first power 
(which therefore vanishes in the low energy limit) and 
to the anomalous magnetic moment of $\Theta^+$.  

In practice, we need to consider the form factor due to the finite 
size of the nucleon.  
Here we adopt a gauge invariant form factor with 
a four momentum cutoff~\cite{ffgauge}.  
This form factor suppresses the nucleon pole
contributions in the PS scheme 
(and hence the contact term also in the PV scheme), 
as reflecting the fact that the nucleon intermediate state is 
far off-shell.  
Consequently, the dominant contribution is given by the 
t-channel process of the kaon exchange and/or $K^*$ meson 
exchange.  
The ambiguity of the anomalous magnetic moment of $\Theta^+$ is 
also not relevant.  
Therefore the difference between the PS and PV schemes 
is significantly suppressed when kaon exchange term  
is present as for the case of the neutron target.  
This allows one to make rather unambiguous theoretical 
predictions.  

We have computed the photoproduction of $\Theta^+$ for the neutron
and proton target, and for the two parities of $\Theta^+$.  
Here are several results:
(1) When the decay width $\Gamma_{\Theta^+ \to KN} = 15$ MeV 
is used the typical total cross section values are 
about 100 [nb] for the positive parity  
and about 10 [nb] for the negative parity.  
Since these values are proportional to $\Gamma_{\Theta^+ \to KN}$, 
experimental information on the decay width is 
useful to determine the size of cross sections. 
In general the cross sections are about ten times 
larger for the positive parity $\Theta^+$ than for the negative 
parity. 
The p-wave coupling $\vec \sigma \cdot \vec q$ effectively enhances the 
coupling strength by factor 3 -- 4 as compared with the s-wave 
coupling for the negative parity,   
when the momentum transfer amounts to 1 GeV.  
(2) For the neutron target, the kaon exchange term is dominant.  
In this case, the $K^*$ contributions are 
not important even with a large $K^*N\Theta$ coupling
$|g_{K^*N\Theta}| = \sqrt{3} |g_{KN\Theta}|$~\cite{close}.  
Hence the theoretical prediction for the neutron target 
is relatively stable.  
The angular dependence has a peak at $\theta \sim 60$ degrees 
in the center-of-mass system, a consequence of the vertex 
structure of the $\gamma KK$ vertex in the kaon exchange term.  
Since this feature is common to both parities,  
the difference in the parity of $\Theta^+$ may not be 
observed in the angular distribution.  
(3) The kaon exchange term vanishes for the case of the proton 
target.  
Therefore, the amplitude is a coherent sum of various Born terms, 
where the role of the $K^*$ exchange is also important.  
The theoretical prediction for the proton target 
is therefore rather difficult.  

\vspace*{0.5cm}

\subsection{Polarized proton beam and target}

This reaction was considered in order to determine the parity of 
$\Theta^+$ unambiguously, independent of reaction 
mechanism~\cite{thomas,nam_pp}.  
In the photoproduction case, the determination of parity is 
also possible if we are able to control the polarization of both 
the initial and the final states~\cite{nakayama}, 
which is however very difficult in the present experimental setup.  

The system of two protons provides a selection rule due to 
Fermi statistics.  
Since the isospin is $I = 1$, the spin and angular momentum of the 
initial state must be either 
$(S, L) = (0, {\rm even})$ or $(S, L) = (1, {\rm odd})$.  
Now consider the reaction 
\be
\vec p + \vec p \to \Theta^+ + \Sigma^+ \, .
\ee
at the threshold region, where the relative motion in the 
final state is in s-wave.  
If the initial spin state has
$S= 0$, then the parity of the final state 
is positive and hence the parity of $\Theta^+$ MUST BE
positive.  
Likewise, if $S=1$ the parity of $\Theta^+$ MUST BE
negative.   

\begin{figure}[t]
\centerline{\includegraphics[width=8cm]
                            {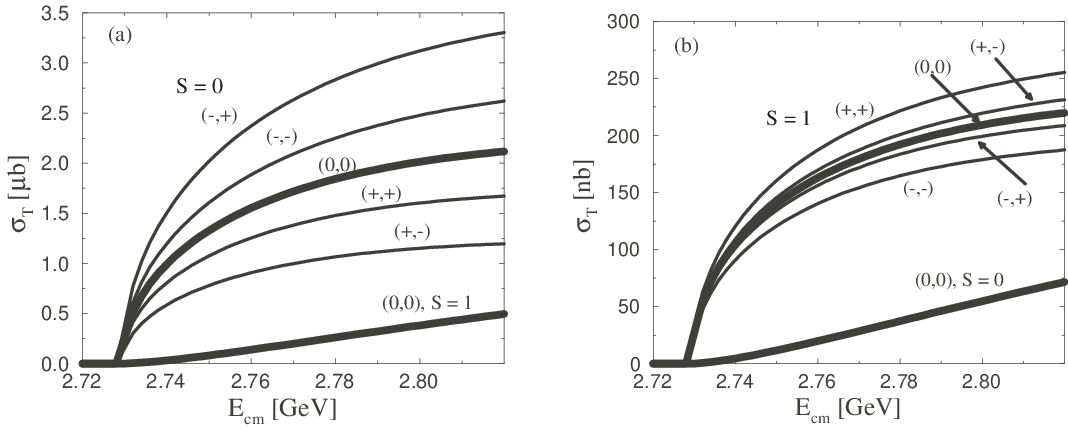}}
\centering
\caption{\small 
Total cross sections for $pp \to \Theta^+ \Sigma^+$, 
(a) for the positive parity and (b) for the negative parity, 
as functions of center of mass energy.  
Different curves correspond to different unknown 
coupling $K^*N\Theta$~\cite{Nam:2003uf}.}
\label{sigma_ppTS}
\end{figure}

One can compute production cross sections by employing an
effective lagrangian of the kaon and $K^*$ exchange model.  
The results are shown in Fig.~\ref{sigma_ppTS} for both 
positive and negative parity $\Theta^+$, where the 
Nijmegen potential for $K$, $K^*$ exchanges~\cite{stokes} and  
decay width 15 MeV are employed.  
The above selection rule is shown clearly by 
the energy dependence at the threshold region. 

Recently COSY-TOF reported the result for the 
$\Theta^+ \Sigma^+$ production in the unpolarized 
$pp$ scattering at $p_p = 2.95$ GeV/c~\cite{cosy_tof}.  
They quote the total cross section 
$\sigma \sim 0.4 \pm 0.2$ [$\mu$b] at 30 Mev above the 
threshold in the center of mass energy.  
In comparison with theory, if we adopt a 
narrower width of about 5 MeV, the cross section 
will be about 0.5 [$\mu$b] for the positive parity 
and 0.05 [$\mu$b] for the negative parity.  
This comparison seems to favor the positive parity $\Theta^+$. 
Although there remain some ambiguities in theoretical 
calculations, such a comparison of the total cross section 
will be useful to distinguish the parity of $\Theta^+$.  

\subsection*{Acknowledgements}

The author would like to thank K.~Hicks, T.~Hyodo, 
H.C.~Kim, T.~Nakano, S.I.~Nam, M.~Oka, E.~Oset, H.~Toki
and A.W.~Thomas for fruitful discussions.


\begin{thebibliography}{10}

\bibitem{Nakano:2003qx}
T.~Nakano {\it et al.}  [LEPS Collaboration],
Phys.\ Rev.\ Lett.\  {\bf 91}, 012002 (2003); 
For the latest experimental situation, see the web site of 
the workshop PENTAQUARK04, www.rcnp.osaka-u.ac.jp/~penta04.  
\bibitem{oka}
For a recent review, see for instance: 
M. Oka, hep-ph/0406211 and references therein.  
\bibitem{quark}
N. Isgur and G. Karl, D {\bf 19}, 2653 (1979); 
ibid. D {\bf 20}, 1191 (1979).
\bibitem{skyrme}
T. H. R. Skyrme, Nucl. Phys. {\bf 31}, 556 (1962); 
G.S. Adkins C.R. Nappi and E. Witten, Nucl. Phys. B 
{\bf 228}, 552 (1983).  
\bibitem{Diakonov:1997mm}
D.~Diakonov, V.~Petrov and M.~V.~Polyakov,
Z.\ Phys.\ A {\bf 359}, 305 (1997) 
\bibitem{cbag}
A. Hosaka and H. Toki, Phys. Reports {\bf 277}, 65 (1996), and 
references therein.  
\bibitem{Hosaka:2003jv}
A.~Hosaka,
Phys.\ Lett.\ B {\bf 571}, 55 (2003).
\bibitem{jennings}
B.K. Jennings and K. Maltman, Phys. Rev. D {\bf 68}, 094020 (2004).
\bibitem{stancu}
Fl. Stancu, D.O. Riska, Phys. Lett. B 573, 242 (2003); 
Fl. Stancu, Phys. Lett. B {\bf 595}, 269 (2004).
\bibitem{carlson}
C.E. Carlson, C.D. Carone, H.J. Kwee and V. Nazaryan, 
Phys. Lett. B {\bf 573}, 101 (2003); hep-ph/1012325.
\bibitem{hosaka2}
A. Hosaka, T. Shinozaki and M. Oka,  in preparation.  
\bibitem{Jaffe:2003sg}
R.~L.~Jaffe and F.~Wilczek,
Phys.\ Rev.\ Lett.\  {\bf 91}, 232003 (2003).
\bibitem{Nam:2003uf}
S.~I.~Nam, A.~Hosaka and H.~C.~Kim,
Phys.\ Lett.\ B {\bf 579}, 43  (2004). 
\bibitem{refphoto}
W.~Liu and C.~M.~Ko, Phys. Rev. C 68, 045203 (2003); 
Q.~Zhao, Phys. Rev. D {\bf 69}, 053009 (2004);
B.~G.~Yu, T.~K.~Choi and C.~R.~Ji, 
nucl-th/0312075, nucl-th/0408006; 
F.E. Close and Q. Zhao, Phys. Lett. B {\bf 590}, 176 (2004); 
Q. Zhao and J.S. Al-Khalili, Phys. Lett. B {\bf 585}, 91 (2004).  
\bibitem{ffgauge}
K. Ohta, Phys. Rev. C {\bf 40}, 1335 (1989); 
H. Haberzettl, C. Bennhold, T. Mart and T. Feuster, 
Phys. Rev. C {\bf 58}, 40 (1998); 
R. M. Davidson and R. Workman, arXiv:nucl-th/0101066.
\bibitem{close}
F.E. Close and J.J. Dudek, Phys. Lett. B {\bf 586}, 75 (2004).
\bibitem{thomas}
A.W. Thomas, K. Hicks and A. Hosaka, 
Prog. Theor. Phys. {\bf 111}, 291 (2004).  
\bibitem{nam_pp}
S.~I.~Nam, A.~Hosaka and H.~C.~Kim, hep-ph/0401074.  
\bibitem{stokes}
V. G. J. Stokes and Th. A. Rijken, Phys. Rev. C {\bf 59}, 3009 (1999).
\bibitem{nakayama}
K. Nakayama and W.G. Love, Phys. Rev. C {\bf 70}, 012201 (2004).  
\bibitem{cosy_tof}
W.~Eyrich et.al., COSY-TOF collaboration, 
Phys. Lett. B {\bf 595}, 127 (2004).  
\end{thebibliography}
\end{document}